\documentclass{PoS}

\title{Early black-hole seeds in the first billion years}

\ShortTitle{BH seeds}

\author{Umberto Maio\thanks{Corresponding author.}\\   % \speaker{Umberto Maio}\\
	INAF -- Italian National Institute for Astrophysics (Italy) \\
	E-mail: \email{ umberto.maio @ inaf.it}}

\abstract{
Supermassive black holes with billion solar masses are in place already within the first Gyr, however, their origin and growth in such a short lapse of time is extremely challenging to understand.
Here, we discuss the formation paths of early black-hole seeds, showing the limits of light black-hole seeds from stellar origin and the expected characteristics of heavy/massive black-hole seeds originated by gas direct collapse in peculiar primordial conditions.
To draw conclusions on the possible candidates and the role of the ambient medium, we use results from N-body hydrodynamic simulations including atomic and molecular non-equilibrium abundance calculations, cooling, star formation, feedback mechanisms, stellar evolution, metal spreading of several heavy elements from SNII, AGB and SNIa, and multifrequency radiative transfer over 150 frequencies coupled to chemistry and SED emission for popII-I and popIII stellar sources.
Standard stellar-origin light black holes are unlikely to be reliable seeds of early supermassive black holes, because, under realistic assumptions, they cannot grow significantly in less than a billion years.
Alternatively, massive black-hole seeds might originate from direct collapse of pristine gas in primordial quiescent mini-haloes that are exposed to stellar radiation from nearby star forming regions.
The necessary conditions required to form these heavy seeds must be complemented with information on the complex features of local environments and the fine balance between chemistry evolution and radiative transfer.
}

\FullConference{
}

\begin{document}

\section{Introduction}

\begin{figure}
\centering
\includegraphics[width=0.245\textwidth]{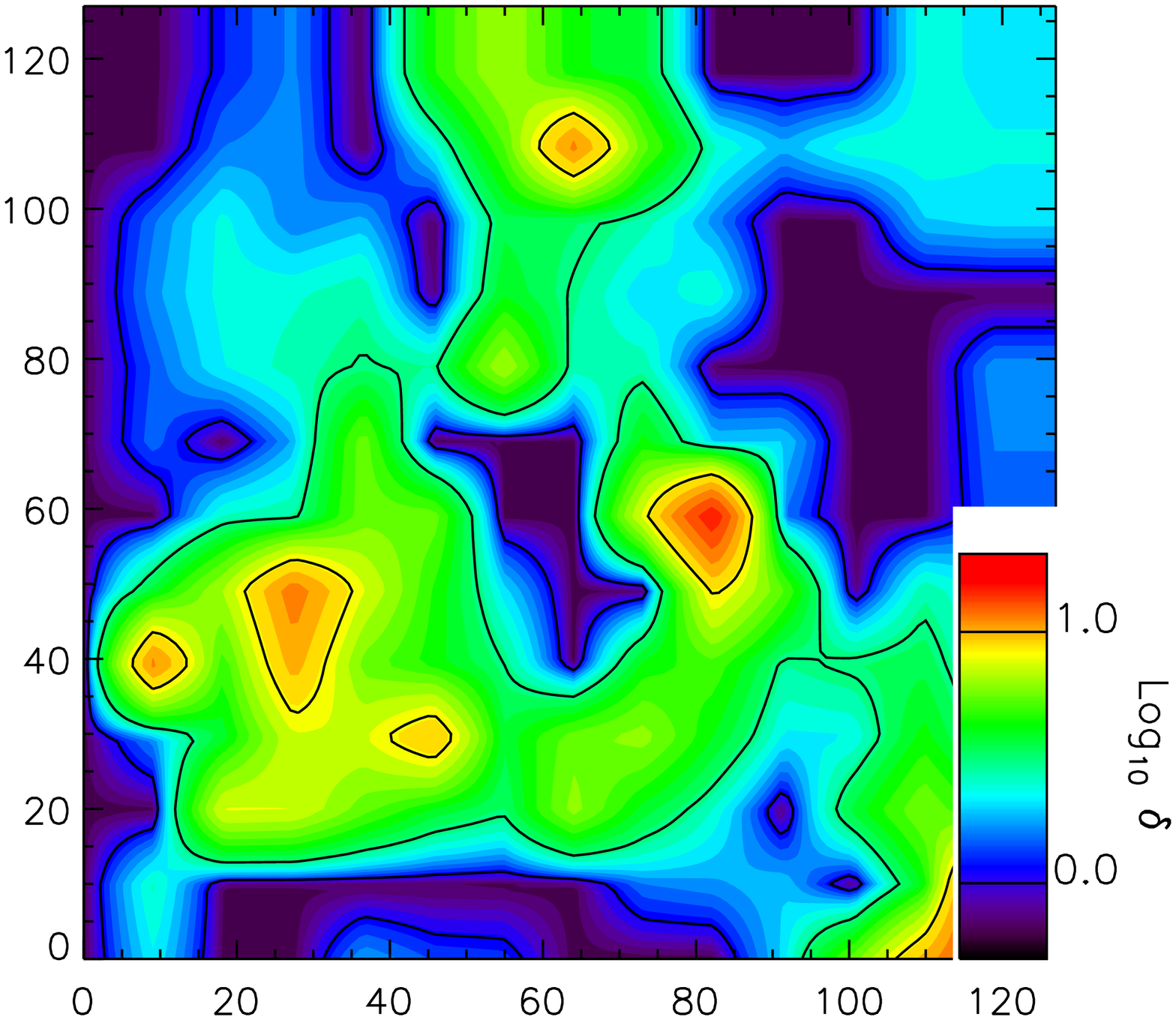}
\includegraphics[width=0.245\textwidth]{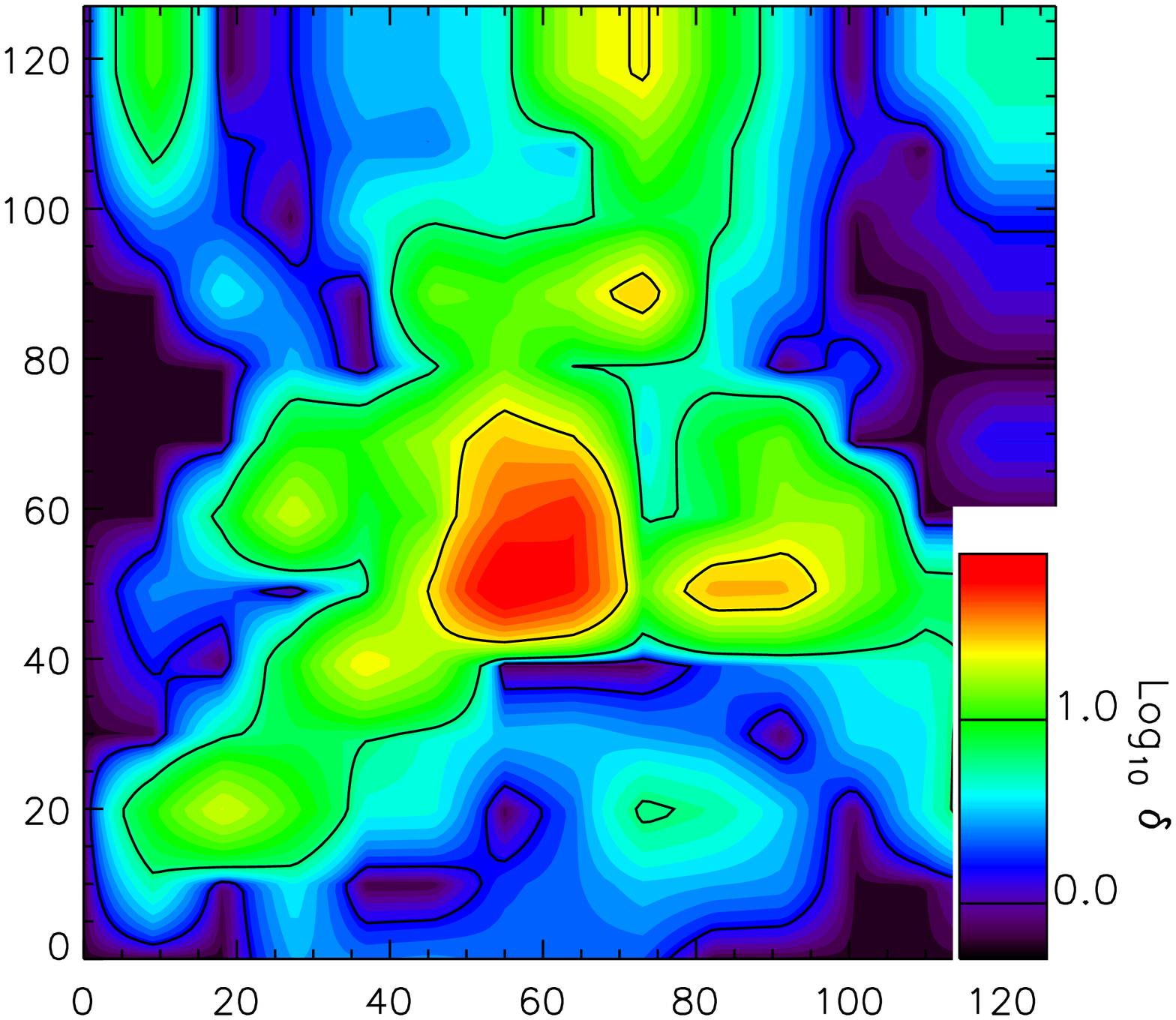}
\includegraphics[width=0.245\textwidth]{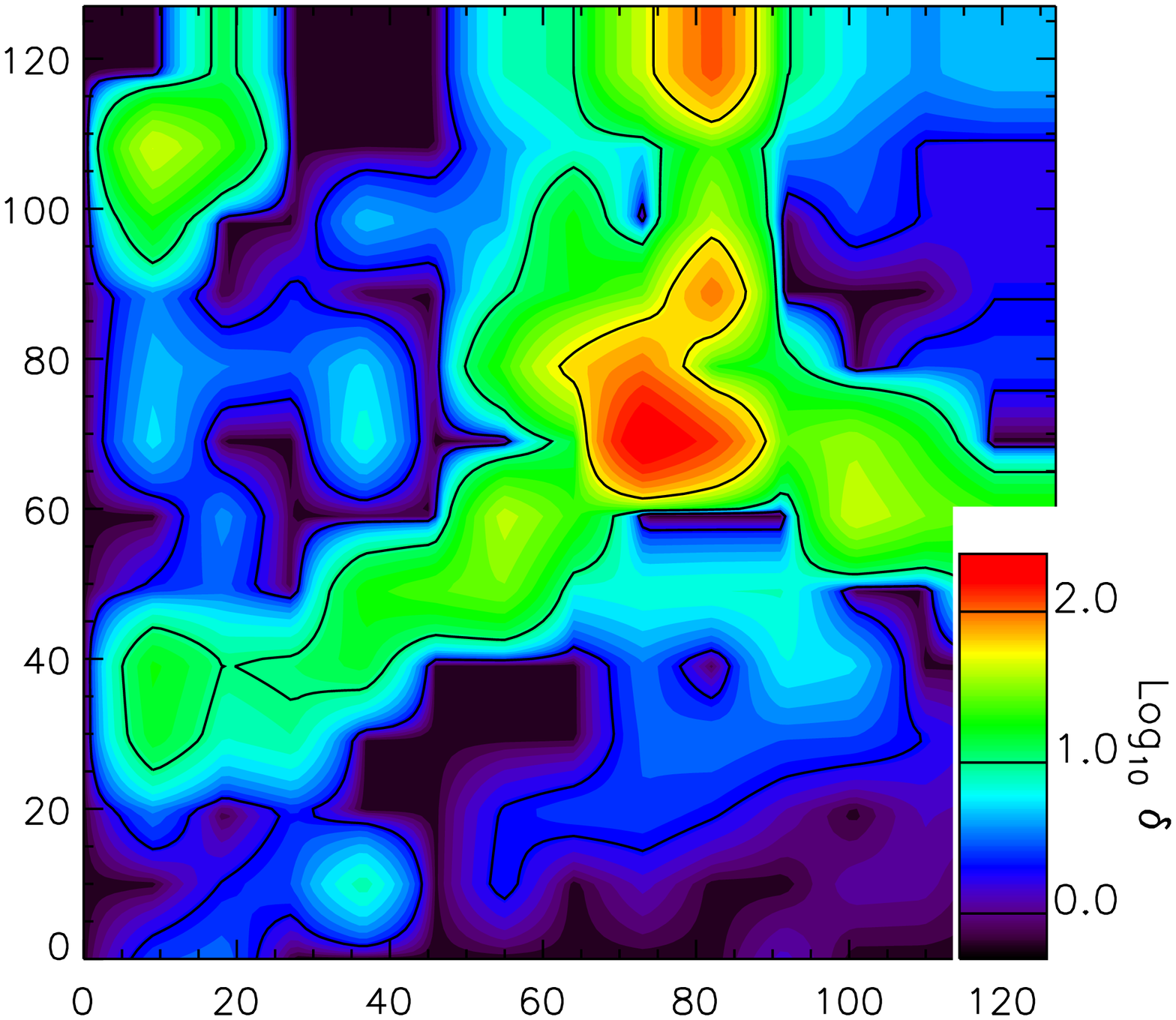}
\includegraphics[width=0.245\textwidth]{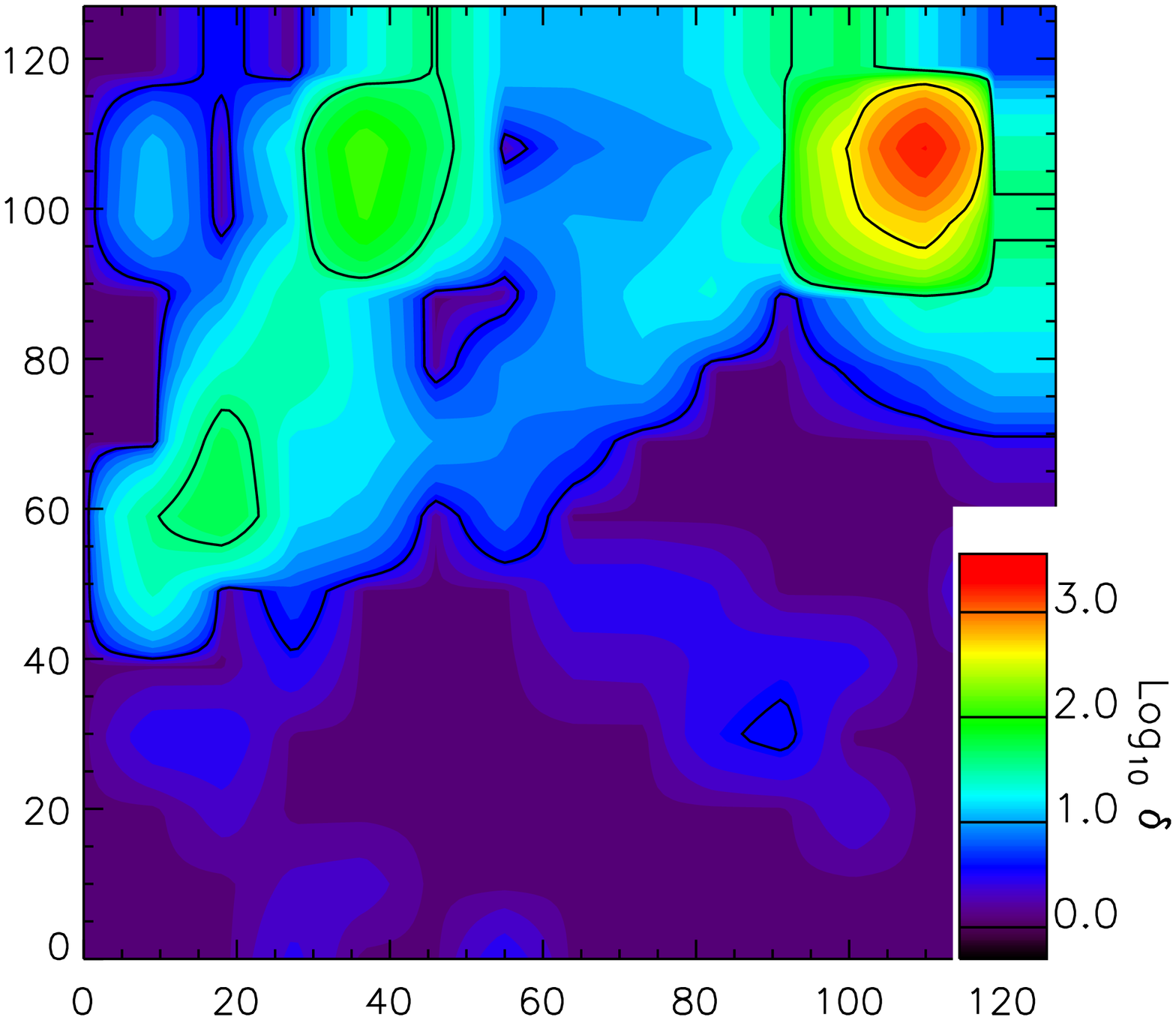}
\caption[]{\small
	Simulated H$_2$-driven gas collapse and inflow of a primordial star forming halo displayed through the projection of gas overdensity, $\delta$, on a 128$\times$128 pixel grid. The region corresponds to $600 \,\rm kpc$ (comoving) at redshift $z \simeq 7$ 
	%	5, 4 and 3 
and the following 0.4, 0.8 and 1.4  Gyr, respectively, from left to right.}
\label{fig:evolution1}
\end{figure}

\begin{figure}
\centering
\includegraphics[width=0.245\textwidth]{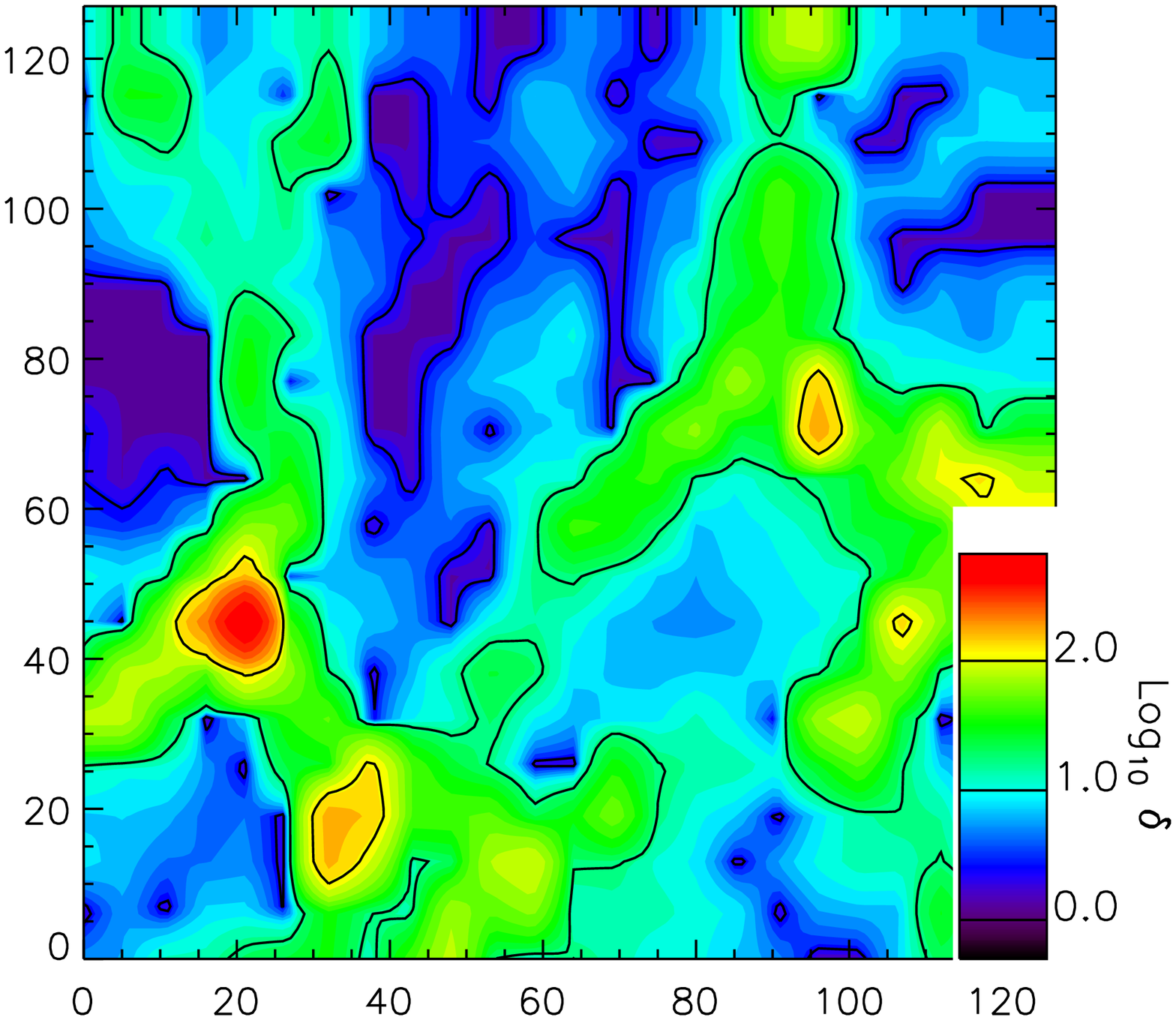}
\includegraphics[width=0.245\textwidth]{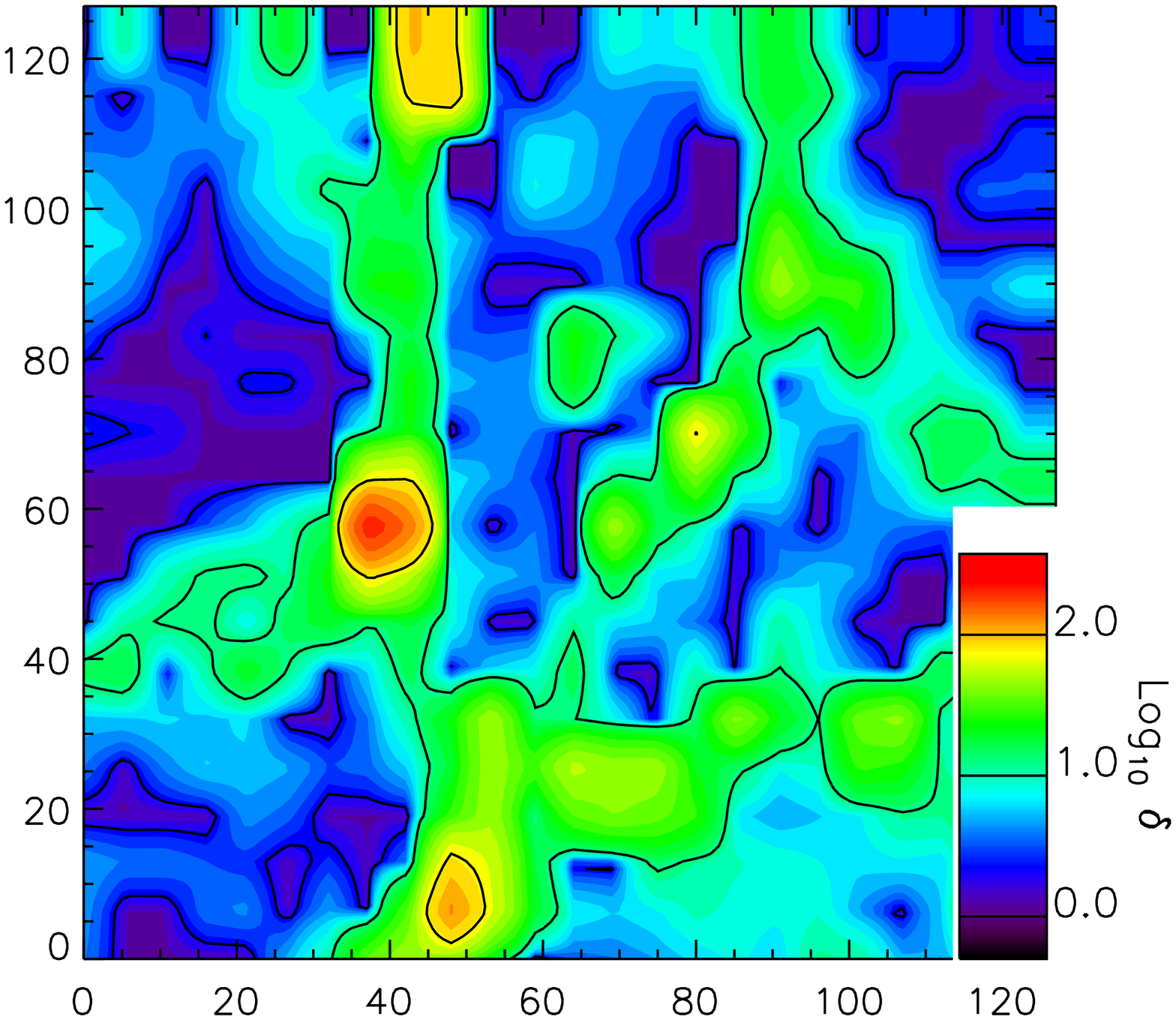}
\includegraphics[width=0.245\textwidth]{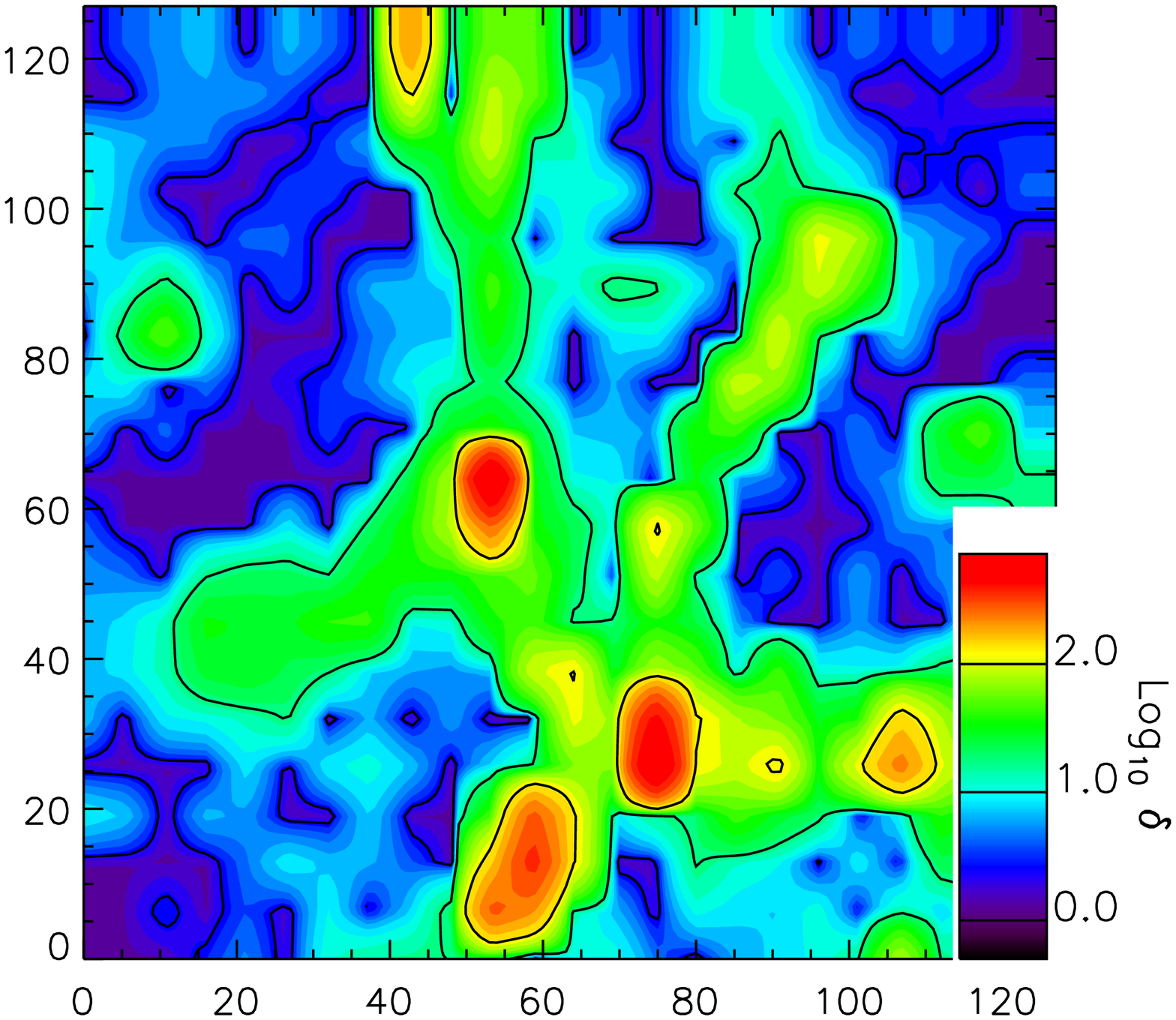}
\includegraphics[width=0.245\textwidth]{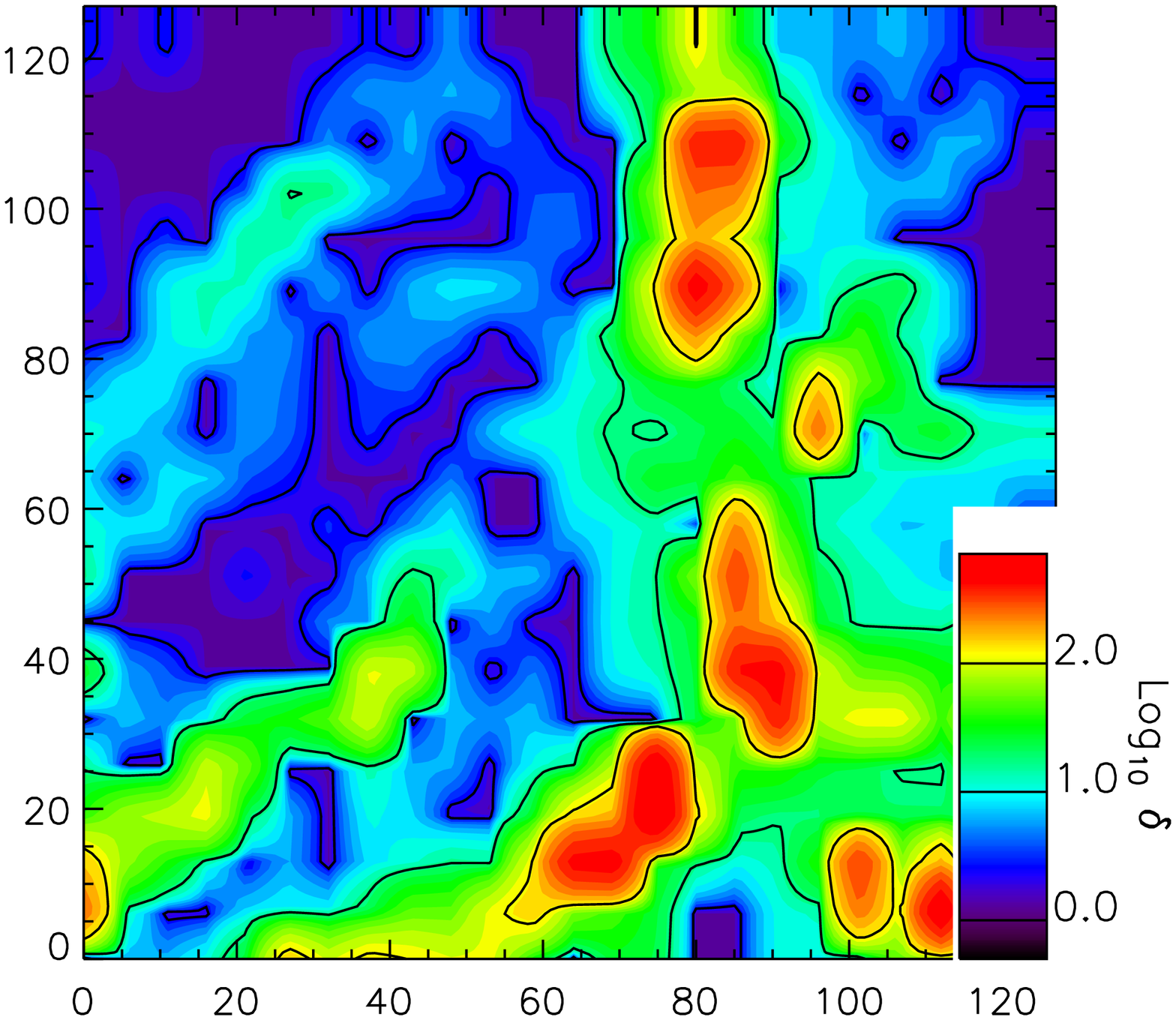}
\caption[]{\small
	Simulated star formation and stellar-origin BH generation in a primordial star forming halo displayed through the projection of gas overdensity, $\delta$, on a 128$\times$128 pixel grid. The region corresponds to $1000 \,\rm kpc$ (comoving) at $z \simeq 7$ over 1.4 Gyr time evolution, as in Fig.~\ref{fig:evolution1}.}
\label{fig:evolution2}
\end{figure}

Black holes (BHs) with billion solar masses are observed up to redshifts $z\simeq 7.5$, when the Universe was only about 0.7 Gyr old \cite{Fan2001, Mortlock2011, Banados2018}.
These evidences make the birth of the first supermassive black holes (SMBHs) one of the most striking events in the first billion years.
However, how such large objects could form and grow in such a short lapse of time is extremely challenging to understand.
The role of the environment in which these extreme objects appear is an unknown issue, as well as the role of the first primordial stellar populations injecting UV photons in the early intergalactic medium (IGM).
In fact, it is still debated whether the origin of SMBHs should be linked to `light' stellar-origin BH seeds (with masses $\sim 1$-$10^2\,\rm M_\odot$) or to `heavier' massive BH seeds (with masses $\sim 10^4$-$10^6\,\rm M_\odot$) \cite{Rees1984}.
Furthermore, it is difficult to predict under which physical conditions these seeds could be born, how they relate to nearby structure formation episodes and what the interplay with the ongoing feedback processes is.
\\
The first baryonic structures are supposed to be born in small primordial dark-matter haloes hosting molecular evolution, star formation and metal pollution.
Primordial haloes are expected to be H$_2$-rich, hence the pristine gas they host can cool and fragment, producing the first population III (popIII) stars, the first stellar black holes and the first heavy elements in cosmic history.
Metal spreading is then responsible for increasing the cooling capabilities of nearby regions \cite{Maio2007}, while the emitted UV photons can ionise the IGM totally or partially depending on stellar masses and explosion energies.
Given that stellar populations described by different initial mass functions (IMFs) produce different amounts of BH remnants, metal spreading and ionizing photons, the actual IMF in star forming regions at different epochs influence significantly BH formation and growth.
Indeed, stellar populations described by conventional Salpeter-like IMFs will produce stellar BHs mostly in the mass range $1$-$10\,\rm M_\odot$ after a few $10^7\,\rm yr$ from their formation time.
Top-heavy IMFs, instead, will lead to the formation of more massive stars and BHs with masses of the order of $\sim 10^2\,\rm M_\odot$.
Metal enrichment is also going to change drastically depending on the original IMF, with typical yields ranging from a few per cent in the Salpeter-like case up to $50\%$ in the top-heavy case.
Furthermore, the emitted photons will create a background radiation field that might ionise and/or dissociate H$_2$ molecules around star forming regions.
The strength of radiation as well as the shape and extension of the impacted regions will depend on the stellar mass range expected by different IMFs.
This is particularly important when addressing the effects of stellar evolution on the chemical and thermal conditions of the cosmic medium. Indeed, pristine gas in early mini-haloes illuminated by UV radiation might keep an atomic state around $10^4\,\rm K$, be prevented from cooling and eventually collapse directly into a massive BH, forming a so-called direct-collapse BH (DCBH).
\\
To shed light on the seeds of SMBHs, it is crucial to study thermal, chemical and radiative properties of cosmic structures during cosmological epochs.
For a complete picture, it is necessary to follow gravity and hydrodynamics coupled to molecule formation and metal production from stellar evolution.
Indeed, (i) molecules lead the first gas collapsing events in primordial pristine gas; 
(ii) metals, spread out by the first stars, increase gas cooling capabilities with their large variety of atomic transitions influencing subsequent structure formation;
(iii) stellar evolution determines typical timescales, metal yields, photon production and BH masses.
Given the high non-linearity of these processes, to address adequately the problem, one has to implement atomic and molecular cooling, star formation, metal enrichment from type II and type Ia supernovae (SNe), as well as AGB phases and radiative transfer (RT) in dedicated N-body hydrodynamic numerical simulations of the first Gyr.
\\
As an example, in Fig.~\ref{fig:evolution1}, simulated H$_2$-driven gas collapse and inflow from a zoom into a primordial star forming halo, as expected by numerical simulations \cite{MT2015}, is displayed. The colour code corresponds to different values of gas overdensity, $\delta$, during a time span of roughly 1.4 Gyr.
In this period, $\delta$ increases from values around unity or less up to $\delta > 10^3$ (red spot in the rightmost panel), where most of the gas mass results concentrated. In typical situations, the overdense gas clump just formed is going to cool, fragment and form stars.
This is highlighted by Fig.~\ref{fig:evolution2} which shows how, simultaneously in the same time span, early gas clumps host first episodes of star formation accompanied by shock heating and metal enrichment in the neighbouring regions. Highly non-linear effects related to the underlying structure evolution (e.g. development of knots and filaments) are evident.
In these chaotic environments, where star formation, feedback and photon production coexist, the first BHs, seeds of the observed SMBHs, can be born.
\\
In the next, we will focus on the possible channels by which early BH seeds form and on the effects from the local environment.
The work is organized as follows: in Sect.~\ref{sect:lightseeds}, we present the main features of light BH seeds, in Sect.~\ref{sect:heavyseeds}, we show the path leading to the formation of heavy BH seeds, while in Sect.~\ref{sect:conclusions}, we summarise and conclude.


\section{Light seeds: stellar black holes} \label{sect:lightseeds}
\begin{figure}[t]
\centering
\includegraphics[width=0.85\textwidth]{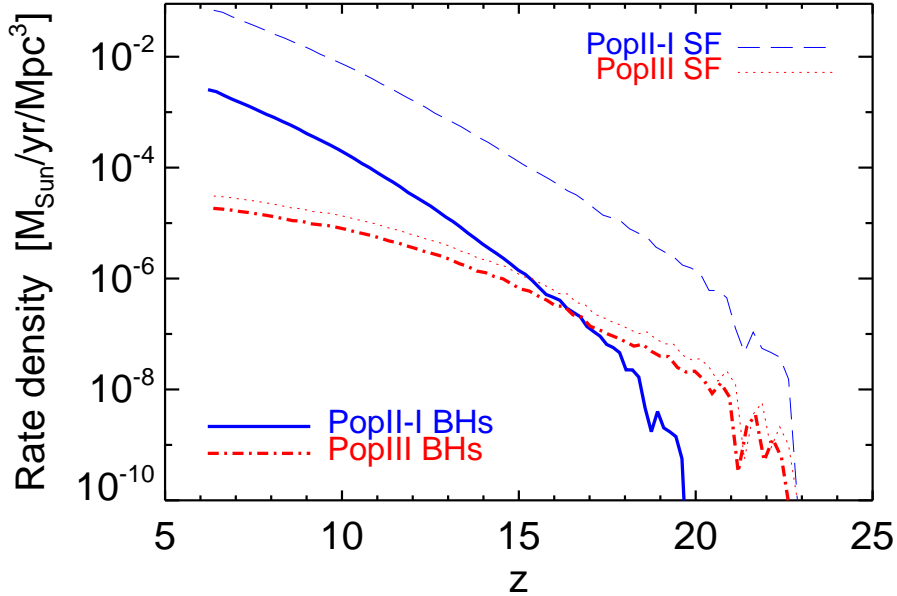}
\caption[]{\small
	Stellar-origin BHs for different populations, as expected from the primordial cosmic star formation rate. The BH formation rate density for popII-I (solid line) and popIII (dash-dotted line) is plotted, as derived from simulated SFR densities for popII-I (dashed line) and popIII (dotted line) regimes.
	}
\label{fig:BHsbyPop}
\end{figure}

Light black-hole seeds are originated at the end of the life of stars.
In early regimes, cosmic gas has pristine or very metal-poor chemical composition and is able to host popIII star formation. 
The typical properties of the popIII regime, such as stellar IMF, explosion energies or BH remnant masses, are still largely unknown.
Theoretical predictions for primordial massive stars suggest that they should produce significant amounts of heavy elements that quickly pollute nearby gas and mark the transition to the standard population II-I (popII-I) regime \cite{Maio2010}.
\\
The popII-I regime takes place in enriched star formation sites (above a critical metallicity of $\sim 10^{-4}\, \rm Z_\odot$) and the expected mass distribution features a power-law shape (Salpeter IMF or similar).
Standard stellar-evolution calculations suggest SN explosion energies around $\sim 10^{51} \,\rm erg$, which is not sufficient to drive large-scale reionization by $z\simeq 6$, but is enough to produce correct star formation rates (SFRs) and  magnitudes \cite{Dayal2013, Mancini2016}, as well as pollute the cosmic medium with metallicities \cite{Salvaterra2013, Maio2013}, abundance ratios \cite{MaQ2017a, MaQ2017b} and dust masses \cite{Mancini2015} consistent with the observations.
The interplay among feedback processes from different stellar populations might affect pollution of the cosmic medium, although details are tightly dependent on the assumptions.
Both popIII and popII-I generations can produce BH remnants.
In the former case, massive popIII stars can leave BHs with masses of the order of $\sim 10^2\,\rm M_\odot$, while in the latter case, one expects remnant BH masses around $\sim 1-10 \,\rm M_\odot$.
This means that stellar-origin BHs are good candidates for light black-hole seeds at all times.\\
Since, the transition from popIII to popII-I generations is quite fast at redshift larger than $\sim 15$, the typical stellar-origin BH masses in the first Gyr are expected to be dominated by popII-I objects.
This is highlighted by Fig.~\ref{fig:BHsbyPop}, where the formation rate of stellar-origin BHs is plotted, as derived from simulated SFR densities for popIII and popII generations.
These numerical calculations are performed with an updated version of the N-body smoothed-particle hydrodynamics (SPH) code Gadget-3 \cite{Springel2005} and the numerical implementation takes into account: primordial non-equilibrium chemistry, atomic and molecular cooling in the temperature regime $\sim 10$-$10^9\,\rm K$ \cite{Maio2007}, star formation, feedback effects and stellar evolution with metal spreading \cite{Tornatore2007} from SNII \cite{WW1995}, AGB \cite{vdHG1997} and SNIa \cite{Thielemann2003} phases for both popIII and popII-I generations according to metal-dependent yields and a critical metallicity for popIII to popII-I transition of $Z_{\rm crit} = 10^{-4}\,\rm Z_\odot$.
The IMF used for regular popII-I regimes is  a Salpeter one over [0.1, 100]~$\rm M_\odot$ range, while the IMF used for popIII regimes is a top-heavy one over [100, 500]~$\rm M_\odot$ range with the same slope as the Salpeter \cite{Maio2010}.
Stars with masses lower than $\sim 8\,\rm M_\odot$ evolve through AGB and SNIa phases, while larger stars evolve as SNII, leaving as remnant a neutron star (if the progenitor mass is lower than $\sim 40\,\rm M_\odot$) or a BH (if the progenitor mass is in the range $\sim 40$-$100~\rm M_\odot$).\footnote{
The threshold mass to distinguish the formation of neutron stars from BHs bears uncertainties with stellar models predicting values between 20~$\,\rm M_\odot$ and 40~$\,\rm M_\odot$ \cite[and references therein]{Campisi2011}. For a Salpeter IMF, the resulting difference on the BH mass fraction is less than a factor of 2 and ranging from 0.037 (with threshold mass of 40~$\,\rm M_\odot$) to 0.074 (with threshold mass of 20~$\,\rm M_\odot$).
% BH mass fraction = 0.0369949 for 40 Msun
% BH mass fraction = 0.0740168 for 20 Msun
}
In this latter case, the expected stellar lifetime is $\lesssim 30 \,\rm Myr$.
PopIII BHs are formed from the death of stars with masses in the ranges 100-140~$\rm M_\odot$ and 260-500~$\rm M_\odot$ after a rapid evolution of roughly 2 Myr. Stars in the intermediate mass range of 140-260~$\rm M_\odot$ die as pair-instability SNe (PISNe) without leaving any remnant.\\
PopII-I SNe have explosion energies of $10^{51}\,\rm erg$ \cite[]{WW1995} and popIII PISNe produce mass-dependent energies between $10^{51}$ and $10^{53}\,\rm erg$ \cite[]{HW2002}.
These events heat, pollute and inject entropy in the surrounding medium.
The metals ejected during stellar evolution are diffused to the neighbouring regions via the SPH kernel, while kinetic feedback is modelled though wind velocities of $500\,\rm km/s$.\\
In our specific case (Fig.~\ref{fig:BHsbyPop}), we adopt cosmological initial conditions at redshift 100 and evolve them down to redshift 6, in a box with side of 10 Mpc/{\it h} with expansion parameter normalised to 100~km/s/Mpc $h=0.7$, present-day matter and $\Lambda$ density parameters $\Omega_{0,m}=0.27$ and $\Omega_{0,\Lambda}=0.73$, flat cosmology and initial number particles of $2 \times 512^3$ for gas and dark matter, respectively \cite[and references therein for more details]{MV2015, MT2015}.
The plot displays popII-I and popIII SFR densities at $z=5$-$25$ and the corresponding popII-I and popIII BH formation rate densities, computed including the appropriate time delays for each population. The trends of the SFR densities are consistent with the ones expected in literature and feature an increasing trend in time, during which popIII star formation (SF, dotted line) becomes more and more subdominant with respect to the popII-I SF (dashed line). Although at $z > 20$ the two regimes are comparable, by redshift $z\simeq 10$ the discrepancy reaches 3 orders of magnitude and keeps increasing.
The corresponding BH formation rates are dictated by the SFR. For the popIII BH formation rate density (dash-dotted line), there is a short time delay induced by the short lifetimes of very massive stars (2 Myr). On the contrary, for popII-I BHs (solid line), the time delay is larger (30 Myr) and this makes the curve steeper than the SFR, mostly at higher $z$, when the effect is more evident due to the non-linearity of the time-redshift relation.
As a result, the solid line increases slightly after $z\simeq 20$ and then intersects the dash-dotted line around $z\simeq 16$. After this redshift, popII-I BHs are produced more abundantly than popIII BHs (by 1 dex around $z\simeq 10$ and 2 dex around $z\simeq 6$).
The epochs before $z\simeq 16$, instead, feature a more significant contribution of popIII BHs (3 dex at $z\simeq 20$), as a consequence of their almost instantaneous link to popIII SFR.
\\
In fact, popII-I BHs dominates at most times, since already the earliest popIII star formation episodes pollute the surrounding haloes above the critical level \cite{Biffi2013} and the popIII contribution to the cosmic SFR density drops irrespectively of exact numerical parameters.
This means that the residual fraction of $\sim 10^2 \,\rm M_\odot$ BH remnants is tiny and available BH seeds lie mostly in the lower mass tail, i.e. $\sim 1$-$10 \,\rm M_\odot$.
Unfortunately, these masses are too small to justify the existence of SMBHs at $z\simeq 7$.
It must be noted that, according to the ordinate values, a hypothetical rapid merger of all the light BHs formed by $z\sim 6$ (in one Gyr) would produce about $10^7 \,\rm M_\odot/Mpc^3$.
Studies investigating accretion processes onto stellar-origin BHs \cite{Hirano2014} find that, even in the most favourable conditions, BH seeds of the order of $\sim 10^2\,\rm M_\odot$ hardly grow in a few hundred Myr.
These considerations imply the need to search for alternative ways to form heavier seed BHs that might grow into SMBHs by $z\simeq 7$ \cite[and references therein]{Valiante2018, Natarajan2019}.

\section{Heavy seeds: direct-collapse black holes} \label{sect:heavyseeds}
\begin{figure}[t]
\centering
\includegraphics[width=0.49\textwidth]{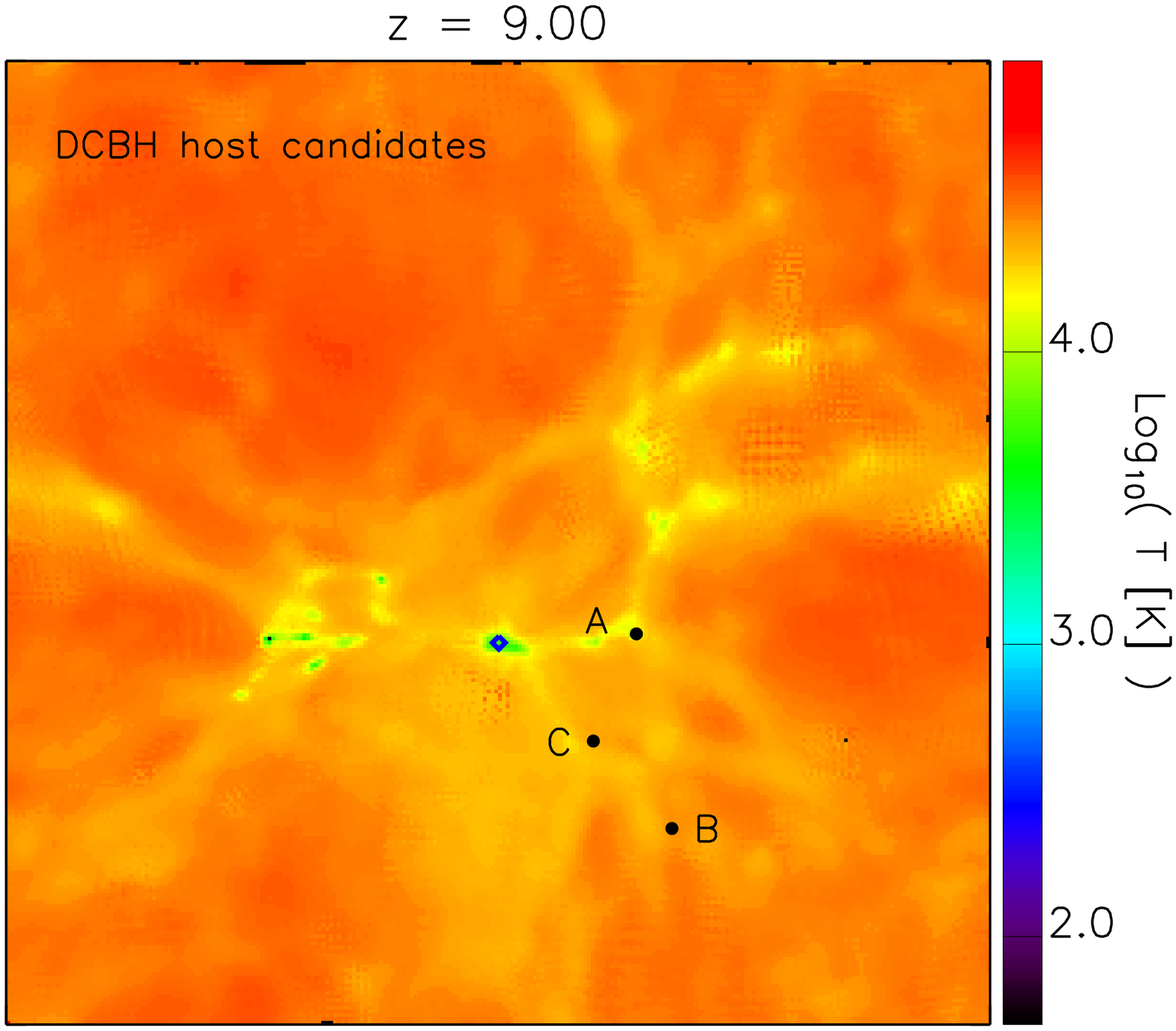}
\includegraphics[width=0.49\textwidth]{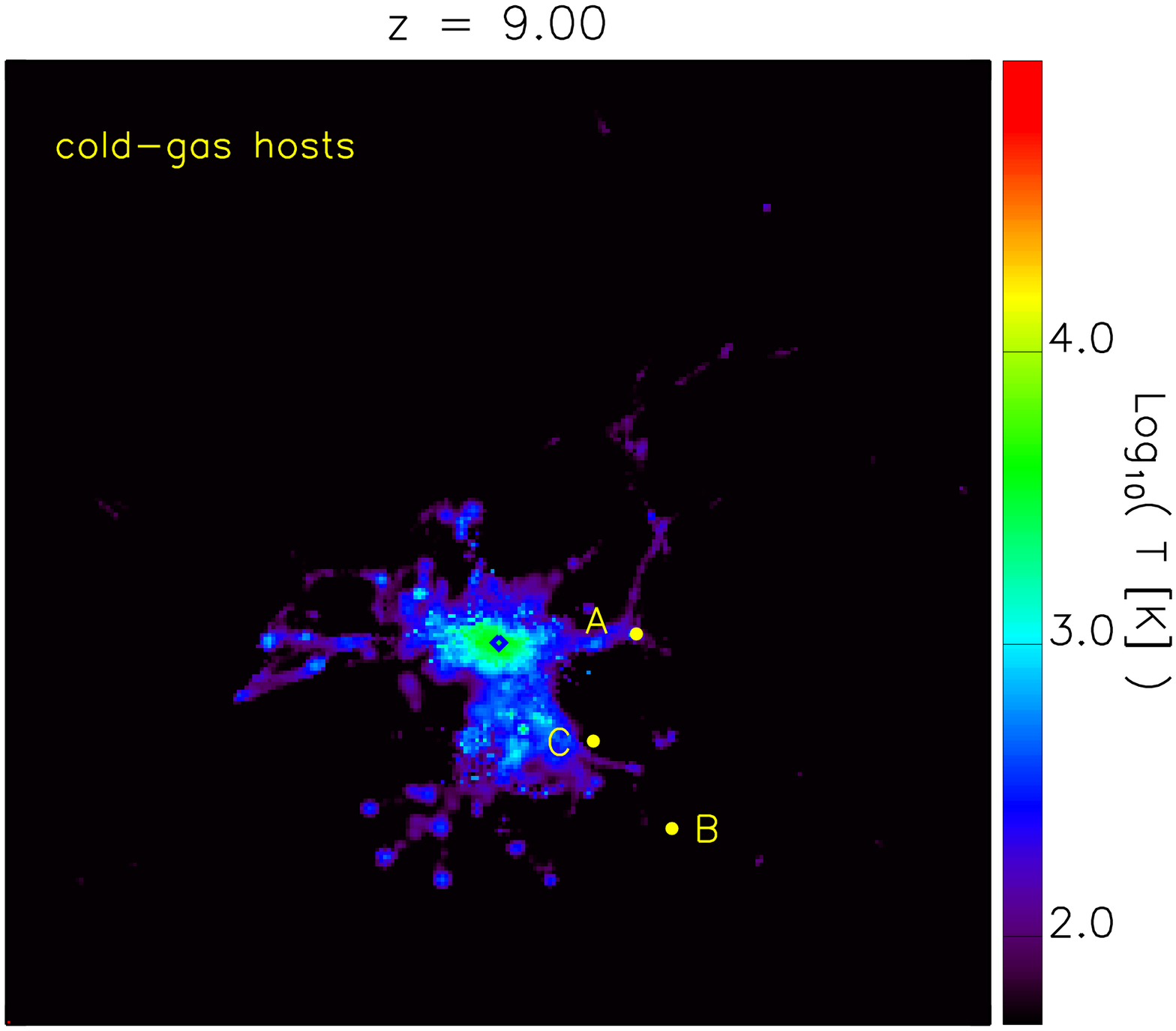}
\caption[]{\small
Temperature maps at $z=9$ for a run with a top-heavy IMF and a powerful SED corresponding to $10^5\,\rm K$ black-body emission (left) and for a run with a Salpeter IMF and a weaker SED corresponding  to $10^4\,\rm K$ black-body emission (right). The black bullets on the left map highlight the positions of DCBH host candidates along cosmic filaments. The yellow bullets on the right map are the same objects, but in this case host cold gas. In both maps, the central star forming region is denoted by an empty diamond.
	}
\label{fig:DCBHhosts}
\end{figure}

A possible alternative to form SMBHs at high redshift is represented by gas accretion on BH seeds with initial masses of $\sim 10^4$-$10^6\,\rm M_\odot$. Forming such heavy seeds is rather difficult, though.
Indeed, dynamical interactions and coalescence of light BHs have long timescales to allow the formation of a single $ 10^4$-$10^6\,\rm M_\odot$ BH in less than a Gyr.
On the other hand, the direct-collapse scenario, a rapid collapse of the gas in primordial haloes into massive BHs \cite{BL2003, Begelman2006}, seems plausible in early epochs.
DCBHs are expected to form in particular environments where {\it pristine} gas is not fragmenting nor forming stars.
This requires that molecular cooling, effective at temperatures below $\sim 8 \times 10^3\,\rm K$, should get dissociated by {\it external UV radiation} in the Lyman-Werner (LW) band.
In order to keep the gas bound in early mini-haloes, one expects that the minimum dark-matter mass of the hosting candidate should be of the order of, or larger than, $\sim 2\times 10^6\,\rm M_\odot$ to prevent photo-evaporation due to photo-heating and feedback mechanisms from nearby star forming regions.
In fact, hosting candidates should be haloes with virial temperature around $10^4\,\rm K$ and exposed to significative amounts of LW flux, $J_{\rm LW}$.
The amount of photons emitted in the LW band depends on the adopted spectral properties of primordial sources \cite{Sugimura2014}, however, literature studies have shown that dissociating LW fluxes (in units of $10^{-21}\,\rm erg/s/cm^2/Hz/sr$, hereafter J$_{21}$) are effective in the range of values $ J_{\rm LW} \sim 1$-$1000\, J_{21}$, while the effects are dramatic for much bigger values, preventing DCBH formation by the end of the first Gyr \cite[etc.]{Shang2010, WG2011,Habouzit2016, JD2017}.
As an example, one might expect that the gas clump in the rightmost panel of Fig.~\ref{fig:evolution1}, where the overdensity is $\delta > 10^3$, under these peculiar conditions might be a good candidate for hosting DCBH formation at redshift $z\gtrsim 10$.
\\
In general, simplified numerical simulations are not capable to address the process of DCBH formation, because the dominant physical mechanism ruling DCBH formation is radiative feedback coupled to non-equilibrium chemistry \cite{PM2012}.
Furthermore, one has to take into account the different spectral energy distributions (SEDs) emitted by sources from different stellar populations to address photo-ionization and photo-heating of the gas in the haloes near star formation sites \cite{Maio2016}.
Indeed, these latter haloes are the ones that might be exposed to sufficient ionizing radiation and be host candidates of DCBHs.
\\
We implement the needed numerical schemes, including all the physical treatments mentioned in the previous section (primordial non-equilibrium chemistry, atomic and molecular cooling, feedback effects, star formation, stellar evolution and metal spreading from SNII, AGB and SNIa for popIII and popII-I generations, a critical metallicity of $Z_{\rm crit} = 10^{-4}\,\rm Z_\odot$) and, in addition, multifrequency RT according to the Eddington tensor formalism coupled to atomic and molecular non-equilibrium calculations for both popII-I and popIII stellar generations \cite{Maio2019}.
The SEDs used are black bodies sampled in the same frequency range as the non-equilibrium chemical calculations (150 frequency bins), in order to compute chemical radiative rates consistently with radiative fluxes from active sources.
For popII-I generations a Salpeter IMF and a popII-I SED approximated as $10^4\,\rm K$ black-body emission are adopted.
For popIII generations we consider two cases: a powerful case with a top-heavy IMF and popIII SED given by a $10^5\,\rm K$ black-body emission; a weak case with popIII stars similar to popII-I, having a Salpeter IMF and a $10^4\,\rm K$ black-body SED.
We adopt initial conditions at redshift 100 and evolve them down to redshift 6, in a box of 0.5 Mpc/{\it h} a side sampled by $2 \times 256^3$ for gas and dark matter, respectively, with $h=0.7$, $\Omega_{0,m}=0.3$ and $\Omega_{0,\Lambda}=0.7$.
These choices are motivated by the need of a trade-off configuration in which physical, chemical and radiative processes are followed accurately while still resolving primordial structures \cite[and references therein]{Maio2019}.
\\
The effects of the different assumptions for IMFs and SEDs from popIII generations are shown in the maps of Fig.~\ref{fig:DCBHhosts}.
DCBH host candidates are searched according to the above-mentioned requirements, however, only in the powerful case three haloes, located several physical kpc from the central popIII star forming region, fulfil the conditions (candidate A, B and C in the Figure).
In the weaker case, there are no candidates, since the structures that are neither forming stars nor impacted by stellar feedback mechanisms typically host cold gas with temperatures below $10^4\,\rm K$.
\\
The three candidates A, B and C have gas masses of $1$-$ 3 \times 10^5 \,\rm M_\odot$ and are exposed to LW radiation of $J_{\rm LW} \sim 1$-$50 J_{21}$, which completely dissociates H$_2$ molecules.
These values for $J_{\rm LW}$  are consistent with the lower-end of the (rather large) range suggested by literature studies.
The gas residing in these haloes shows turbulent patterns at all resolved scales with Reynolds numbers around $10^4$-$10^8$, consistently with what expected for early mini-haloes at these epochs \cite{Maio2011}.
The distance of the candidates A, B and C from the central source is larger than 5 physical kpc. This value should be considered as a lower limit for DCBH formation, since gas closer to the central star forming region is heavily affected by metal enrichment and stronger photo-evaporation due to the larger LW intensities.
At distances $\gtrsim 5\,\rm kpc$, material is not significantly polluted and photo-evaporation is milder as a consequence of the decease of radiation flux with distance.
Since metal spreading inhibits DCBH formation near star forming regions, where $J_{\rm LW}$ is larger, one must conclude that this channel to form heavy seeds requires a very fine balance between chemical feedback and radiative feedback.
Nevertheless, mechanical feedback due to local substructures or clump interactions may further harm DCBH formation.
A closer look to the three haloes reveals that candidate A is composed by one single roundish structure, candidate B by an irregular elongated shape where gas is interacting and candidate C by two distinct sub-clumps.
Such local non-linear processes will influence the evolution of the latter two candidates at later times and possibly inhibit DCBH formation \cite{Maio2019}, although in the former case a DCBH may still form.
This suggests that, despite DCBHs need extremely peculiar conditions to form, they could explain at least part of the SMBH population at high redshift.

\section{Conclusions}  \label{sect:conclusions}
We have investigated the path to the formation of the early seeds of SMBHs.
To address self-consistently the formation of light and heavy BH seeds in primordial environments, we have developed and analysed numerical hydrodynamic simulations including the relevant physical and chemical processes taking place at early cosmological epochs: atomic and molecular non-equilibrium chemistry, cooling, star formation, feedback mechanisms, stellar evolution for popIII and popII-I generations, metal spreading from SNII, AGB and SNIa, as well as coupling with a multifrequency RT scheme according to adopted SEDs for popII-I and popIII radiative sources.
\\
Light BH seeds originated during the final stages of standard stellar evolution can form within the first billion years with masses of the order of $\sim 1$-$10^2 \,\rm M_\odot$.
Their formation rate is tightly related to the SFR, although delayed by the mass-dependent stellar lifetimes of their progenitors (Fig.~\ref{fig:BHsbyPop}).
The formation rate of light BH seeds at early times ($z\gtrsim 16$) is initially dominated by popIII remnants for three main reasons: 
(i)  popIII SF contribute significantly during the first bursts of cosmic star formation;
(ii) popIII BHs have masses of the order of $10^2\,\rm M_\odot$, i.e. lifetimes of about 2 Myr, so their formation rate follows closely the SFR;
(iii) popII-I BHs have masses of the order of $1$-$10\,\rm M_\odot$, i.e. lifetimes of at least 30 Myr, so their formation rate is delayed with respect to popIII ones.
Later on, at redshift $z\lesssim 16$ popII-I remnants quickly take over, driven by the increasing popII-I SF, and their formation rate density becomes 2 dex larger then the popIII ones by $z \simeq 6$.
Since by then light BH seeds cannot experience significant accretion, their role for SMBH formation is marginal. \\
Given the variety of configurations of their host candidates and the environmental features in which seed BHs might form, it is worth considering seeding mechanisms that require alternative scenarios rather than just focusing on star formation in massive halos \cite{Buchner2019}.
In particular, heavy BH seeds could form through direct collapse of pristine gas in a single BH and give birth to a so-called direct-collapse BH, DCBH.
After careful studies and within the limits of our simulations, we find that only powerful primordial stellar generations can determine the necessary conditions for DCBH formation, while standard stellar populations are too weak and, hence, unable to prevent molecule formation and fragmentation in the ambient gas.
The DCBH host candidates feature gas masses of $1$-$ 3 \times 10^5 \,\rm M_\odot$, turbulent regimes, pristine chemical composition and H$_2$ dissociation due to LW radiation in the range $J_{\rm LW} = 1$-$50 J_{21}$.
They are found at a minimum distance of 5 physical kpc from the irradiating star forming site, being closer regions affected by photo-evaporation and metal enrichment that prevent gas direct collapse.
Local non-linear processes, such as substructure formation and interactions, will influence DCBH evolution, as well. Thus, the necessary conditions required to form DCBHs must be complemented with information on the complex features of the local environment and a very fine balance between chemistry evolution and radiative transfer.\\
Although DCBHs might need extremely peculiar conditions, they could explain at least part of the SMBH population at high redshift.
We note that here we strictly focus on the early phases of BH formation. The processes related to the following accretion and growth are complex issues that lie beyond the scope of this work.
UV background distortions might be present and could be traced by metal ions \cite{Graziani2019}, however their effects at such primordial times should have little impact on cosmic structure evolution.
Alternative scenarios adopted for the cosmological model \cite{Maio2006q}, primordial non-gaussianities \cite{MaioIannuzzi2011, Maio2011ng, MK2012} or dark-matter nature \cite{MV2015} are unlikely to change substantially our conclusions on early BH seeds as derived from the properties of primordial cosmic star formation and metal enrichment \cite{Maio2012ng-grb}.
Primordial supersonic baryon streaming motions have the effect of delaying gas collapse, however, realistic values for supersonic streaming velocities would cause delays of some $10^7\,\rm yr$ \cite{Maio2011vb}. This corresponds to variations in $z$ of a few at very early epochs and smaller at lower $z$, therefore, primordial streaming motions will induce modest delays to BH seed formation, but they will not cause dramatic changes to the whole picture.
The existence of tiny BHs formed during inflation has also been suggested, as well as BH formation from quarks or exotic dark matter.
Given the large extent of such topics, we refer the interested reader to literature studies \cite{Rubin2001, Khlopov2005, Khlopov2010, Belotsky2019}.
\\
In the future, observational synergies \cite[etc.]{Whalen2013, SKA2015} will shed light on primeval epochs giving information about the early state of neutral gas \cite{Hutter2019} and constraining the impacts of different radiative feedback models \cite{Zackrisson2019}. Combinations of optical, IR and X-ray surveys \cite{Haiman2019} will be promising for detecting early seed BHs and discerning between BH seeding channels using electromagnetic observations at redshifts $z>10$.

\section*{Acknowledgments}
We acknowledge detailed comments by the the referee, A. Dolgov, as well as useful discussions with D. Whalen, A. Santangelo and W. Kundt.
This work has been supported through a research grant awarded to U.M. by the German Research Fundation (DFG) project n. 390015701 and the EU HPC-Europa3 Transnational Access Programme project n. HPC17ERW30. Numerical simulations and data analysis have been performed under the PRACE-2IP Programme, grant agreement n. RI-283493, and supported by the facilities of the Italian Computing Center (CINECA) and the Leibniz Institute for Astrophysics, Germany.
We acknowledge the NASA Astrophysics Data System (ADS) and the JSTOR archive for their bibliographic tools.

\end{document}